\pdfoutput=1
\documentclass[twocolumn,10pt,a4paper,byrevtex,floatfix,prl,unsortedaddress,superscriptaddress]{revtex4}
\usepackage[utf8]{inputenc}
\usepackage{microtype}
\usepackage{newtxtext}
\usepackage[upint]{newtxmath}
\usepackage{eucal}

\usepackage{graphicx}
\usepackage{booktabs}
\usepackage{enumerate}
\usepackage{siunitx}
\usepackage{braket}
\usepackage{color}
\usepackage{float}
\usepackage{soul}
\usepackage{bm}

\usepackage[colorlinks,allcolors=blue]{hyperref}
\usepackage[capitalize]{cleveref}

\usepackage[svgnames]{xcolor}
\usepackage{todonotes}
\setuptodonotes{inline}

\definecolor{DarkRed}{rgb}{0.80,0,0}
\definecolor{DarkBlue}{rgb}{0.20,0,0.80}
\definecolor{Purple}{rgb}{0.55,0,0.55}

\begin{document}
\title{Dzyaloshinskii--Moriya-type spin-spin interaction from mixed-parity superconductivity}
\author{Jabir Ali Ouassou}
\affiliation{Department of Computer Science, Electrical Engineering and Mathematical Sciences, Western Norway University of Applied Sciences, NO-5528 Haugesund, Norway}
\author{Takehito Yokoyama}
\affiliation{Department of Physics, Tokyo Institute of Technology, Meguro, Tokyo 152-8551, Japan}
\author{Jacob Linder}
\affiliation{Center for Quantum Spintronics, Department of Physics, Norwegian \\ University of Science and Technology, NO-7491 Trondheim, Norway}

\begin{abstract}
    Interacting impurity spins adsorbed on surfaces have been suggested as basic components for applications in quantum computation and spintronics.
    Such spins usually prefer a parallel or antiparallel configuration, but weakly non-collinear alignments are possible due to the Dzyaloshinskii--Moriya interaction (DMI) that arises in the presence of relativistic spin-orbit coupling.
    Here, we show that an effective Dzyaloshinskii--Moriya-type interaction (DMTI) can emerge purely from superconducting correlations without any spin-orbit interaction.
    We give an analytical proof and provide a numerical study which shows that DMTI arises in mixed-parity superconductors solely from the superconducting pairing.
    Moreover, we show that the same effect can be realized in Josephson junctions between $s$-wave and $p$-wave superconductors, where a phase bias toggles the DMTI entirely on and off.
    These results enable a way to engineer spin textures using superconducting order.
\end{abstract}

\maketitle

\textit{Introduction.}---%
Individual spins placed on the surface of a material, as well as magnetic layers separated by a non-magnetic layer, can communicate with each other over distance via electrons in the separating material.
This is the Ruderman--Kittel--Kasuya--Yosida (RKKY) \cite{ruderman_pr_54, kasuya_ptp_56, yosida_pr_57} interaction and has a dual purpose.
Firstly, it determines the preferred magnetic configuration of spins, which could be used as ``write'' operations in spintronics devices.
The ``read'' operation is performed by measuring the spin configuration.
Secondly, the ground-state spin orientation provides important information about correlations and interactions in the mediating material, acting as a probe for the host environment.
Notably, localized spins interacting via exchange is sufficient to implement a universal quantum computer~\cite{loss_quantum_1998, divincenzo_universal_2000}, so RKKY interactions could possibly provide an architecture for future quantum computers.

\begin{figure}[tb!]
    \includegraphics[width=\columnwidth]{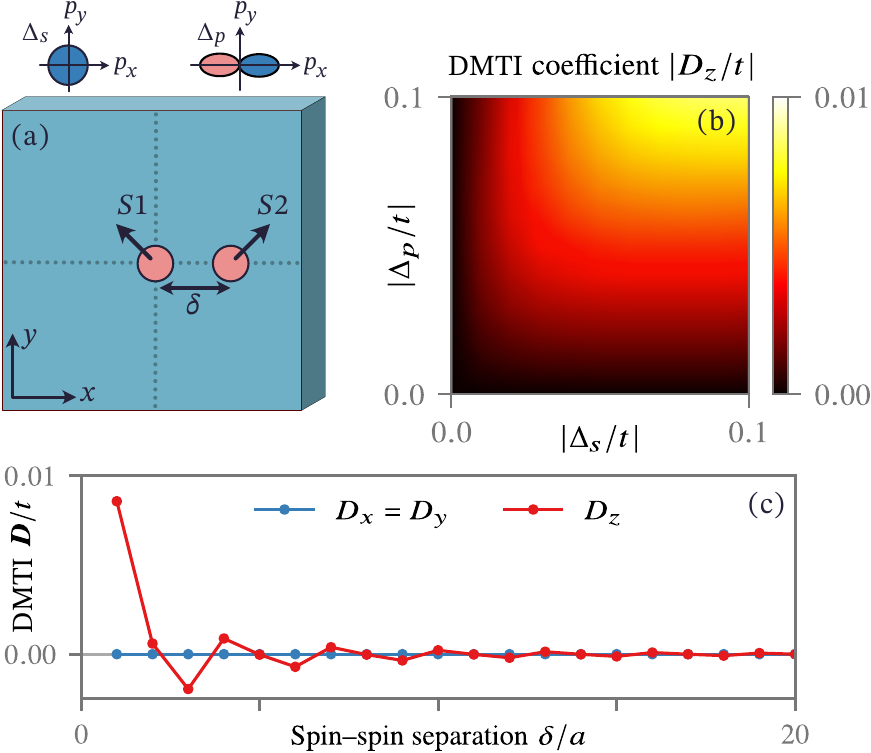}
    \caption{%
        \textbf{(a)}~We consider an $s+ip_x$ superconductor of dimensions $101a \times 101a$ in the $xy$ plane, where $a$ is the lattice constant.
        This superconductor has order parameter contributions $\Delta_s$ and $\Delta_p$ that are respectively symmetric and antisymmetric in momentum space, as illustrated.
        One spin impurity is placed at coordinates $(51a, 51a)$ and another at $(51a+\delta, 51a)$, and their spin orientations couple via an indirect exchange interaction mediated by the superconductor.
        \textbf{(b)}~Dzyaloshinskii--Moriya-type interaction (DMTI) $D_z$ describing the interaction between $\bm{S}_1$ and $\bm{S}_2$ as function of the $s$-wave singlet gap $\Delta_s$ and $p$-wave triplet gap $\Delta_p$ for $\delta = a$.
        Only when both components are non-zero, corresponding to a mixed-parity superconducting state, do we find a finite DMI term.
        There is no SOC in our model, so this DMI term arises purely from mixed-parity superconductivity.
        \textbf{(c)}~DMTI as a function of $\delta$ for $\Delta_s = \Delta_p = 0.1t$, which exhibits the same damped oscillations as usual RKKY interactions.
    }
    \label{fig:intro}
\end{figure}

RKKY interactions have been studied in many material classes, including metals~\cite{yafet_prb_87, imamura2004a}, topological insulators~\cite{shiranzaei_prb_17, zhu_prl_11, liu_prl_09}, and superconductors \cite{alekseevskii_zetf_77, kochelaev_zetf_79, khusainov_zetf_96, aristov_zpb_97, dibernardo2019a, rodriguez2023a}.
In its most basic form, where two spins couple via a normal metal, the spins align either in parallel (P) or antiparallel (AP) depending on their separation distance.
However, spin interactions promoting a non-collinear ground state configuration come into play when adding spin-orbit coupling (SOC).
This manifests through an emergent Dzyaloshinskii--Moriya interaction (DMI) \cite{dzyaloshinsky_jpcs_58, moriya_pr_60} term in the free energy.
Non-collinear spin arrangements are relevant to many subfields of condensed matter physics due to their role in chiral magnetic textures such as skyrmions \cite{nagaosa_natnan_13}, and could also facilitate ferroelectric control over magnetism due to the coupling between DMI and electric polarization \cite{fert_jpsj_23}.

When DMI arises due to SOC, its magnitude is severely limited due to its relativistic origin.
Therefore, an important question is if one can generate strong Dzyaloshinsky--Moriya-type interactions (DMTI) by other means than conventional SOC.
We here predict that this is indeed possible in mixed-parity superconductors with an $s+ip_x$ symmetry [see \cref{fig:intro}(a)].
This kind of superconductivity arises both intrinsically in noncentrosymmetric materials \cite{Smidman2017} and by design in heterostructures consisting of conventional superconductors and ferromagnets~\cite{eschrig_jltp_07}.
Our findings show that such materials provide a new platform for applications that require non-collinear spins and superconductivity, e.g.\ Kitaev chains constructed from magnetic adatoms on superconductors.
For other mixed-parity superconductors, e.g.\ $s+i(p_x+p_y)$, our analytical result shows that DMTI exists for spins displaced along either the $x$ or $y$ axis.
It may then be possible to realize e.g.\ skyrmions in a 2D lattice of magnetic adatoms, stabilized by superconductivity-induced DMTI instead of SOC.

\textit{Theory.}---%
When two spin impurities $\bm{S}_1$ and $\bm{S}_2$ are placed on a superconductor, the free energy of this perturbation can to leading order in $\bm{S}_1$ and $\bm{S}_2$ be parametrized as~\cite{ouassou_rkky_2024}
\begin{equation}
    \mathcal{F} = \mathcal{F}_0 + \bm{\mu} \cdot (\bm{S}_1 + \bm{S}_2) + \bm{J} \cdot (\bm{S}_1 \circ \bm{S}_2) + \bm{D} \cdot (\bm{S}_1 \times \bm{S}_2),
    \label{eq:free-energy}
\end{equation}
where $\mathcal{F}_0$ is independent of the spin directions $\bm{S}_1$ and $\bm{S}_2$, $\bm\mu$ describes a magnetic interaction between each spin and a triplet superconducting condensate, $\bm{J} = (J_x, J_y, J_z)$ describes a Heisenberg- or Ising-type RKKY interaction, $\bm{D} = (D_x, D_y, D_z)$ describes a Dzyaloshinsky--Moriya-type RKKY interaction, and $\circ$ is a Hadamard product.
In Ref.~\cite{ouassou_rkky_2024}, we have previously developed both numerical and analytical methodologies to determine $\bm{J}$ and $\bm{D}$ for general superconductors.
Moreover, we determined these coefficients for pure $s$-wave and $p$-wave superconductors, respectively.
Interestingly, it was possible to obtain a DMTI in non-unitary $p$-wave superconductors without SOC.
However, that interaction was very weak and only occurred for very specific spin placements, since the mechanism requires proximity to the system's edges.

In this paper, we go beyond our previous work by studying a mixed-parity superconductor with an $s+ip_x$ order parameter.
We show that a new DMTI arises in this system which (i)~does not require proximity to system edges, and (ii)~is many orders of magnitude larger than the previously discovered DMTI.
\emph{In fact, this new DMTI becomes larger than all other RKKY coefficients for some parameter ranges, which is important for both experimental validation and potential applications.}

The RKKY interaction energy $\mathcal{E}$ can be analytically obtained via a 2nd-order perturbation expansion in the exchange coupling~$\mathcal{J}$ between each spin and the superconductor underneath.
This perturbation expansion allows us to understand the microscopic origin of the terms in \cref{eq:free-energy}.
The result of that procedure is the standard equation~\cite{ouassou_rkky_2024, schwabe1996a, imamura2004a}
\begin{equation}
    \begin{aligned}
        \mathcal{E} \sim \mathcal{J}^2 \, \mathrm{Im} 
        \int &\mathrm{d}\bm{p}_1 \int \!\mathrm{d}\bm{p}_2 \, \mathrm{e}^{-i(\bm p_2 - \bm p_1) \cdot (\bm R_2 - \bm R_1)}  \int \!\mathrm{d}\omega \, \tanh(\omega/2T) \\
             &\times \mathrm{Tr}\big[ (\bm S_1 \cdot \hat{\bm\sigma}) \hat{G}^R(\bm p_1, \omega) (\bm S_2 \cdot \hat{\bm \sigma}) \hat{G}^R(\bm p_2, \omega) \big],
    \end{aligned}
    \label{eq:rkky-energy}
\end{equation}
where $\bm{S}_1$ and $\bm{S}_2$ are the two spin orientations, $\hat{G}^R$ is the retarded $4\times4$ Green function in Nambu$\otimes$Spin space, $\hat{\bm \sigma} = \mathrm{diag}(\bm\sigma, {\bm\sigma}^*)$ where  $\bm\sigma = (\sigma_1, \sigma_2, \sigma_3)$ is the Pauli vector, $\bm{p}_{1,2}$ are momentum variables, $\omega$ is the quasiparticle energy, and $T$ is the temperature.
We can parametrize a mixed-parity Green function as
\begin{equation}
    \begin{aligned}
        \hat{G}^R = \begin{pmatrix}
            (g_s + \bm{g}_p \cdot \bm\sigma)\sigma_0 & (f_s + \bm{f}_p \cdot \bm\sigma)i\sigma_2 \\
            (\tilde{f}_s + \tilde{\bm{f}}_p \cdot \bm{\sigma}^*) i\sigma_2 & (\tilde{g}_s + \tilde{\bm{g}}_p \cdot \bm{\sigma}^*) \sigma_0 \\
        \end{pmatrix},
    \end{aligned}
    \label{eq:green-param}
\end{equation}
where $\{ g_s, \bm{g}_p, f_s, \bm{f}_p \}$ are functions of momentum~$\bm p$ and energy~$\omega$, $\sigma_0$ is the identity matrix, and tilde conjugation is defined as $\tilde{x}(\bm p, \omega) = x^*(-\bm p, -\omega)$.
Here, $g_s$ and $\bm{g}_p$ are related to the spin-independent and spin-dependent density of states, whereas $f_s$ and $\bm{f}_p$ can be interpreted as $s$-wave singlet and $p$-wave triplet superconducting correlations.
Next, we substitute \cref{eq:green-param} into \cref{eq:rkky-energy} and take the traces.
Finally, we extract the parts of $\mathcal{E}(\bm S_1, \bm S_2)$ that can be written $\bm{D} \cdot (\bm{S}_1 \times \bm{S}_2)$, and thus find the following result for $\bm{D}$:
\begin{widetext}
    \begin{equation}
        \begin{aligned}
            \bm{D} \sim \mathcal{J}^2 \, \int \mathrm{d}\bm{p}_1 & \int \mathrm{d}\bm{p}_2 \, \int \mathrm{d}\omega \, \tanh(\omega/2T) \,  \sin[(\bm{p}_2 - \bm{p}_1) \cdot (\bm{R}_2 - \bm{R}_1)] \\
                                                                 & \times \,\Big\{ \, \mathrm{Re} \big[ \bm{g}_p(\bm{p}_1, \omega) \times \bm{g}_p(\bm{p}_2, \omega) \big] + 2 \, \mathrm{Im}\big[ g_s(\bm{p}_1, \omega) \bm{g}_p(\bm{p}_{2}, \omega) \big] - 2 \, \mathrm{Im}[ f_{s}(\bm{p}_1, \omega) \tilde{\bm{f}}_p(\bm{p}_2, \omega) \big] \,\Big\}.
        \end{aligned}
        \label{eq:analytics}
    \end{equation}
  \end{widetext}
Note that the sine term ensures $\bm{D}(\bm{R}_1, \bm{R}_2) = -\bm{D}(\bm{R}_2, \bm{R}_1)$, as expected of antisymmetric exchange interactions.
We have discarded a DMTI term proportional to $\bm{f}_p \times \tilde{\bm{f}}_p$ which vanishes in bulk systems for symmetry reasons.
The contributions from $\bm{g}_p$ and $g_s$ can arise due to e.g.\ SOC without any superconductivity~\cite{imamura2004a}.
However, and importantly, the $f_s \tilde{\bm{f}}_p$ term can only arise in the presence of both $s$-wave singlet and $p$-wave triplet superconductivity with different complex phases, which is intrinsically realized for $s+ip$ superconductors.
Motivated by this last term, we from here on focus on an $s+ip_x$ mixed-parity superconductor as a host material for spins.

\textit{Numerics.}---%
For most simulations presented herein, we considered a $101a\times101a$ square lattice with open boundary conditions.
One spin was placed at the lattice center [see \cref{fig:intro}(a)] and another spin displaced a distance $\delta$ along the $x$ axis.
By choosing an odd number of lattice sites and placing one spin at the system center, we completely suppress potential edge-induced DMTI contributions~\cite{ouassou_rkky_2024}.
The tight-binding Hamiltonian for this system is
\begin{equation}
    \begin{aligned}
        \mathcal{H} =\; & 
        \mathcal{H}_0
        - \mu \sum_{i\sigma} c^\dagger_{i\sigma} c_{i\sigma}
        - t \sum_{\langle ij \rangle \sigma} c^\dagger_{i\sigma} c_{j\sigma} \\
                        & - \frac{1}{2} \mathcal{J} \sum_{i\sigma\sigma'} \sum_{p=1,2} \delta_{i,i_p} c^\dagger_{i\sigma} (\bm{S}_p \cdot \bm{\sigma})_{\sigma\sigma'} c_{i\sigma'} \\
                        & - \sum_{ij\sigma\sigma'} \left\{ c^\dagger_{i\sigma} \left[ \left( \Delta_{s} \delta_{ij} + \Delta_p \bm{d}_{ij} \cdot \bm{\sigma} \right) i\sigma_2 \right]_{\sigma\sigma'} c^\dagger_{j\sigma'} + \text{h.c.} \right\}.
    \end{aligned}
    \label{eq:hamiltonian-op}
\end{equation}
Here,  $\mathcal{H}_0$ is a constant that is unimportant for our (non-selfconsistent) calculations, and $c^\dagger_{i\sigma}, c_{i\sigma}$ are the usual electronic creation and annihilation operators at lattice site~$i$ for spin~$\sigma$.
Except where otherwise stated, we used a nearest-neighbor hopping $t = 1$, chemical potential $\mu = -3t$, exchange interaction $\mathcal{J} = 3t$, $s$-wave singlet gap $\Delta_s = 0.1t$, and $p$-wave triplet gap $\Delta_p = 0.1t$.
The vector $\bm{d}_{ij}$ is determined from the $d$-vector $\bm{d}(\bm{p}) = i\bm{e}_z p_x$ of the $p$-wave order parameter.
Finally, the lattice sites $i_1$ and $i_2$ correspond to the positions of $\bm{S}_1$ and $\bm{S}_2$, respectively.
\emph{Note that there is \textit{no SOC} in this Hamiltonian, usually a prerequisite for DMTI.}

After numerically constructing the Hamiltonian matrix from \cref{eq:hamiltonian-op}, we calculated the free energy for 36 spin configurations $\bm{S}_1, \bm{S}_2 \in \{ +\bm{e}_x, +\bm{e}_y, +\bm{e}_z, -\bm{e}_x, -\bm{e}_y, -\bm{e}_z \}$ and 20 spin separations $\delta \in [1a, 20a]$.
The free energy is given by
\begin{equation}
    \mathcal{F} = \mathcal{H}_0 - \frac{1}{2} \sum_{\epsilon_n > 0} \epsilon_n - T \sum_{\epsilon_n > 0} \log(1 + \mathrm{e}^{-\epsilon_n/T}),
\end{equation}
where $\{ \epsilon_n \}$ are the numerically calculated eigenvalues and $T = 0.001t$ is the temperature.
The Python code used to construct the tight-binding model and calculate the free energy has been published as free and open-source software~\cite{bodge}.
These calculations are sufficient to fit \cref{eq:free-energy} to the numerical results, thus obtaining $\{ \bm{\mu}(\delta), \bm{J}(\delta), \bm{D}(\delta) \}$.
We found no magnetic interaction between the spins and superconductor ($\bm \mu = 0$), while the RKKY parameters ($\bm{J}$~and~$\bm{D}$) were generally finite.

\textit{Results.}---%
\Cref{fig:intro}(b) shows $D_z(\Delta_s, \Delta_p)$.
This vanishes when either $\Delta_s \to 0$ or $\Delta_p \to 0$, indicating that mixed-parity order is essential for this interaction.
We have also performed calculations for $s+p_x$ superconductors without the `$i$' (not shown), and found no DMTI in those systems.
These findings are consistent with \cref{eq:analytics} having a term $\bm D \sim \mathrm{Im}[f_s \bm{f}_p]$.

\begin{figure}[tb!]
    \centering
    \includegraphics[width=\columnwidth]{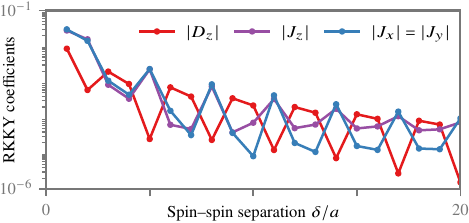}
    \caption{%
        Magnitudes of all finite RKKY coefficients in an $s+ip_x$ superconductor with $\Delta_s = \Delta_p = 0.1t$.
        The DMTI $D_z$ has similar magnitude to the conventional RKKY coefficients $\{ J_x, J_y, J_z \}$, but it dominates strongly for some separation distances~$\delta$.
    }
    \label{fig:comp}
\end{figure}

\Cref{fig:intro}(c) shows $D_z(\delta)$ for an $s+ip_x$-wave superconductor.
Strikingly, the resulting DMTI is over two orders of magnitude larger than the edge-induced contribution in non-unitary triplet superconductors identified in Ref.~\cite{ouassou_rkky_2024} for similar system parameters.
The DMTI oscillates and decays as a function of $\delta$, which entails that very different spin configurations can be expected as a function of the distance between the spins.
As expected from the analytical result in \cref{eq:analytics}, we find that $\bm{D} \sim \mathrm{Im}\,\bm{f}_p^* \sim \mathrm{Im}\,\bm{d} \sim \bm{e}_z$, where we have used that the anomalous Green function $\bm{f}_p$ points in the same direction as the $d$-vector $\bm{d} \sim i\bm{e}_z$.

For an $s+ip_x$ superconductor, a DMTI was only found for spin displacements along the $x$ axis; when spins are displaced purely along the $y$ axis, the DMTI remains zero.
This can be understood from \cref{eq:analytics}.
For $x$-displacements, we have $\bm{R}_2 - \bm{R}_1 = \delta \bm{e}_x$.
For a general $s+ip_x$ superconductor, we can write $f_s(\bm{p}_1, \omega) \tilde{\bm{f}}_p(\bm{p}_2, \omega) = F(|\bm{p}_1|, |\bm{p}_2|, \omega) p_{2x} \bm{e}_z$ for some function $F$.
The integrand is therefore proportional to $p_{2x} \sin[\delta (p_{2x} - p_{1x})]$, which remains finite when integrated over all momentum directions.
For $y$-displacements, one similarly obtains $p_{2x} \sin[\delta (p_{2y} - p_{1y})]$, which vanishes when integrated over positive and negative $p_{2x}$.
For more complicated order parameters (not shown), e.g.\ an $s+ip$ superconductor with the $d$-vector $\bm{d} = (\bm{e}_x + i\bm{e}_y) (p_x + ip_y)$, we find DMTI with different orientations for displacements along the $x$ and $y$ axes---consistent with the arguments above.

The DMTI we obtain is also consistent with Moriya's original symmetry arguments~\cite{moriya_pr_60}.
The simulations presented herein correspond to an $s+ip_x$ superconductor, where the superconducting order parameter transforms as described in the supplemental material of Ref.~\cite{ueno2013}:
Under e.g.\ mirroring through the $xz$ plane, $\Delta(p_x, p_y, p_z) \to -\sigma_2 \Delta(p_x, -p_y, p_z) \sigma_2^{*}$, where $\Delta(\bm{p})$ is the $2\times2$ order parameter in spin space
\begin{equation}
\Delta(\bm{p}) = [\Delta_s + \Delta_p (\bm{d}(\bm{p}) \cdot \bm\sigma)] i\sigma_2.
\end{equation}
Applied to an $s+ip$ superconductor with the $d$-vector $\bm{d} = p_x \bm{e}_z$, we find that the order parameter is invariant under mirroring through the $xy$ and $yz$ planes.
When the spins are displaced along the $x$ axis, the $xy$ plane becomes a mirror plane containing both $\bm{R}_1$ and $\bm{R}_2$, so Moriya's 3rd rule requires that $\bm{D} \parallel \bm{e}_z$.
Moreover, the $yz$ plane is then a mirror plane perpendicular to $\bm{R}_2 - \bm{R}_1$, so Moriya's 2nd rule requires that $\bm{D} \perp \bm{e}_x$.
The numerical result $\bm{D} \sim \bm{e}_z$ for displacements along the $x$ axis is consistent with both constraints.
Contrastingly, if the spins are displaced along the $y$ axis, both the $xy$ and $yz$ planes become mirror planes that include $\bm{R}_1$ and $\bm{R}_2$.
Thus, Moriya's 3rd rule requires that $\bm{D} \parallel \bm{e}_z$ and $\bm{D} \parallel \bm{e}_x$, which is only possible for $\bm{D} = 0$.
This result for displacements along the $y$ axis is thus required by symmetry.

In \cref{fig:comp}, we compare the order of magnitudes of the non-zero RKKY coefficients for the above system.
Since all coefficients oscillate as functions of~$\delta$, the dominant RKKY contribution depends on the exact separation distance.
For instance, when $\delta = 6a$ or $\delta = 9a$, the DMTI is roughly an order of magnitude larger than the Heisenberg contribution.
On the other hand, for e.g.\ $\delta = 5a$ the Heisenberg contribution is roughly two orders of magnitude larger than the DMTI.
The non-DMTI contribution is roughly Heisenberg-type ($J_x = J_y = J_z$) for small separation distances, but acquires an increasing Ising contribution ($J_z \neq J_x = J_y$) as the spins are moved further apart.
Thus, an $s+ip_x$ superconductor with spin impurities provides a rich set of ground-state spin configurations: depending on the precise placements of each spin, the dominant RKKY interaction can be Heisenberg-type, Ising-type, or Dzyaloshinsky--Moriya-type.

\begin{figure}
    \includegraphics[width=\columnwidth]{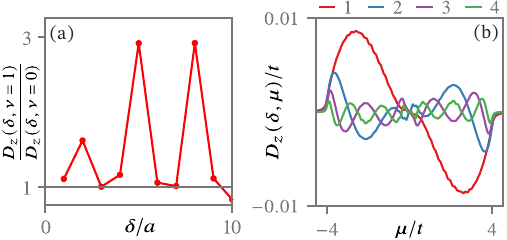}
    \caption{%
        Tunability of the DMTI in a mixed-parity superconductor.
        \textbf{(a)}~Relative change in $D_z$ when the phase winding~$\nu$ becomes finite.
        \textbf{(b)}~$D_z$ as a function of the chemical potential~$\mu$, for different $\delta/a$ as indicated in the legends above.
        For this specific plot, we used a $51a\times51a$ lattice size to make the numerics feasible.
    }
    \label{fig:tuning}
\end{figure}

In \cref{fig:tuning}(a), we compute the influence of a supercurrent flowing through the superconductor on the RKKY interaction \cite{sun_prb_24} to determine if the DMTI can be experimentally tuned.
To this end, we include a phase winding $\Delta_{s,p} \rightarrow \Delta_{s,p} \mathrm{e}^{2\pi i \nu x/L}$, where $x$ is the position along the $x$ axis and $L$ is the length of the system.
Thus, $\nu = 0$ corresponds to no phase winding (the case studied so far in this manuscript), while $\nu = 1$ corresponds to a $2\pi$ phase winding across the sample.
To ensure current conservation along the $x$ axis, periodic boundary conditions were used for these specific simulations.
We see that the DMTI can be increased by up to 192~\% for certain~$\delta$.
The largest modulation is found for $\delta = 5a$ and $\delta = 8a$, corresponding to near-zero values of $D_z$ without phase winding [cf.\ \cref{fig:intro}(c)].
This permits \emph{in situ} engineering of the DMTI between the spins.
Note that the applied current in most cases \emph{enhances} the DMTI---only for $\delta=10a$ do we find a weak reduction of $D_z$.
The same results were obtained whether the current is applied in the positive or negative direction ($\nu = \pm 1$).

In \cref{fig:tuning}(b), we show the DMTI as a function of the chemical potential~$\mu$.
Interestingly, the DMTI exhibits an oscillating dependence on $\mu$ which is antisymmetric with respect to $\mu=0$, and decays exponentially for $|\mu| > 4t$.
When $\delta$ increases, the wavelength and amplitude of these oscillations decrease.
The main features of this behavior can be understood from \cref{eq:analytics}.
As discussed above, the DMTI integrand contains a factor $p_{2x} \sin[\delta(p_{2x} - p_{1x})]$ in $s+ip_x$ superconductors.
The momentum factor $p_{2x} - p_{1x}$ is constrained by the Fermi momentum in the system, which in turn is a function of the Fermi level~$\mu$.
This sine function thus explains why the DMTI oscillates both as a function of $\delta$ and $\mu$.
The fact that the DMTI vanishes for $|\mu| > 4t$ can be understood because the RKKY interaction is mediated by correlations at the Fermi level.
When $|\mu| > 4t$, the density of states is shifted so much that no electronic states remain at the Fermi level.
In reality, the decay near $|\mu| \approx 4t$ is likely to be more abrupt, since we might expect the superconducting gap itself to vanish in this limit, which is not captured by our non-selfconsistent simulations.
The results in \cref{fig:tuning}(b) suggest an alternative way to tune the DMTI \emph{in situ}: In 2D superconductors, the chemical potential can to some extent be tuned via a gate voltage, permitting voltage control over the RKKY DMTI.

\begin{figure}[tb]
    \includegraphics[width=\columnwidth]{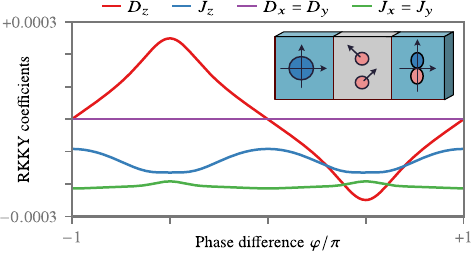}
    \caption{%
        Plot of the RKKY interactions in the normal metal interlayer of a Josephson junction as a function of the phase difference across the junction. We here consider one superconductor to be a conventional $s$-wave superconductor, and the other to be a $p$-wave superconductor with the $d$-vector $\bm{d} = p_y \bm{e}_z$. The DMTI dominates when the phase difference $\varphi \to \pm \pi / 2$. Inset: Illustration of the setup considered. The junction has dimensions $61a\times41a$, where each superconductor has a length $20a$ and the interlayer has length $21a$. We then placed one spin at the exact center of the system, and displaced the other by $\delta = 4a$ along the $y$-axis.}
    \label{fig:phase}
\end{figure}

In \cref{fig:phase}, we consider a different physical realization of the effect: using a Josephson junction constructed from one $s$-wave and one $p_y$-wave superconductor separated by a normal metal interlayer. When the phase difference is tuned to $\pm \pi/2$, we create an effective $s+ip_y$ mixed-parity superconducting state in the proximitized interlayer, which results in a DMTI there. This shows that our results can be used as a way to probe experimentally whether a candidate material might be a $p$-wave superconductor, by connecting it to a conventional superconductor in a Josephson junction and checking for a phase-induced DMTI. The phase difference across a Josephson junction can be tuned using an applied magnetic flux or an injected electric current.

One way to measure  RKKY interactions is spin-polarized scanning tunneling microscopy (SP-STM) as demonstrated in Ref.~\cite{zhou_natphys_10}.
When spin impurities are placed on a surface as in \cref{fig:intro}(a), the electric current tunneling between a spin and SP-STM tip depends on their relative spin orientations: The current is maximized for parallel magnetizations and minimized for antiparallel magnetizations.
From the $\mathrm{d}I/\mathrm{d}V$ curve as a function of applied magnetic field, one can determine to what extent the spins are parallel or antiparallel.
In the case of a strong DMTI, we would of course expect the ground state configuration to be neither parallel nor antiparallel but rather somewhere in-between.
A similar approach was used to study the RKKY DMTI specifically in Ref.~\cite{schmitt_indirect_2019}, where the SP-STM results show oscillations as a function of position along an atomic chain.
Their system is particularly relevant for a chain of spin impurities placed on an $s+ip_x$ superconductor, as discussed with respect to Kitaev chains earlier in this manuscript.

\textit{Concluding remarks.}---%
In this paper, we have shown analytically and numerically that a Dzyaloshinsky--Moriya-type contribution to the RKKY interaction arises specifically in $s+ip$ mixed-parity superconductors.
Moreover, we have shown that this DMTI becomes the dominant spin--spin interaction for some parameter ranges, and that its precise magnitude can be tuned \emph{in situ} either via a charge supercurrent or via a gate voltage.
Notably, these effects are obtained without any spin-orbit coupling in our model.
This suggests mixed-parity superconductors as a promising platform for engineering non-collinear magnetic ground states for artificial spin chains and lattices.

\begin{acknowledgments}
    This work was supported by the Research Council of Norway through Grant No.\ 323766 and its Centres of Excellence funding scheme Grant No.\ 262633 ``QuSpin.''
    This work was also supported by JSPS KAKENHI Grant Number JP30578216.
    The numerical calculations were partially performed on resources provided by Sigma2---the National Infrastructure for High Performance Computing and Data Storage in Norway, project NN9577K.
    The research presented in this paper has also benefited from the Experimental Infrastructure for Exploration of Exascale Computing (eX3), which is financially supported by the Research Council of Norway under contract 270053.
\end{acknowledgments}

\newpage
\bibliography{references}
\end{document}